\documentclass[a4paper,12pt]{article}

\usepackage{latexsym}
\usepackage[small,hang,bf]{caption} 

\newcommand{\nc}{\newcommand}    

\nc{\be}[1]{\begin{equation}\mbox{$\label{#1}$}}
\nc{\bea}[1]{\begin{eqnarray} \mbox{$\label{#1}$}}
\nc{\Section}[2]{\section{#2}\label{#1}}
\nc{\Bibitem}[1]{\bibitem{#1}}
\nc{\Label}[1]{\label{#1}}

\nc{\eea}{\end{eqnarray}}
\nc{\ee}{\end{equation}}

\nc{\bdm}{\begin{displaymath}}
\nc{\edm}{\end{displaymath}}
\nc{\dpsty}{\displaystyle}
\nc{\bc}{\begin{center}}
\nc{\ec}{\end{center}}
\nc{\ba}{\begin{array}}
\nc{\ea}{\end{array}}
\nc{\bab}{\begin{abstract}}
\nc{\eab}{\end{abstract}}
\nc{\btab}{\begin{tabular}}
\nc{\etab}{\end{tabular}}
\nc{\bit}{\begin{itemize}}
\nc{\eit}{\end{itemize}}
\nc{\ben}{\begin{enumerate}}
\nc{\een}{\end{enumerate}}
\nc{\bfig}{\begin{figure}}
\nc{\efig}{\end{figure}}

\nc{\arreq}{&\!=\!&}
\nc{\arrmi}{&\!-\!&}
\nc{\arrpl}{&\!+\!&}
\nc{\arrap}{&\!\!\!\approx\!\!\!&}
\nc{\non}{\nonumber}
\nc{\align}{\!\!\!\!\!\!\!\!&&}

\def\lsim{\; \raise0.3ex\hbox{$<$\kern-0.75em
      \raise-1.1ex\hbox{$\sim$}}\; }
\def\gsim{\; \raise0.3ex\hbox{$>$\kern-0.75em
      \raise-1.1ex\hbox{$\sim$}}\; }

\nc{\DOT}{\hspace{-0.08in}{\bf .}\hspace{0.1in}}
\nc{\Laada}{\hbox {$\sqcap$ \kern -1em $\sqcup$}}
\nc\loota{{\scriptstyle\sqcap\kern-0.55em\hbox{$\scriptstyle\sqcup$}}}
\nc\Loota{{\sqcap\kern-0.65em\hbox{$\sqcup$}}}
\nc\laada{\Loota}
\nc{\qed}{\hskip 3em \hbox{\BOX} \vskip 2ex}

\nc{\real}{{\rm I \! R}}
\nc{\Z}{{\sf Z \!\!\! Z}}
\nc{\complex}{{\rm C\!\!\! {\sf I}\,\,}}
\def\bigid{\leavevmode\hbox{\small1\kern-3.8pt\normalsize1}}
\def\id{\leavevmode\hbox{\small1\kern-3.3pt\normalsize1}}
\nc{\slask}{\!\!\!/}
\nc{\bis}{{\prime\prime}}
\nc{\pa}{\partial}
\nc{\na}{\nabla}
\nc{\ra}{\rangle}
\nc{\la}{\langle}
\nc{\goto}{\rightarrow}
\nc{\swap}{\leftrightarrow}

\nc{\EE}[1]{ \mbox{$\cdot10^{#1}$} }
\nc{\abs}[1]{\left|#1\right|}
\nc{\at}[2]{\left.#1\right|_{#2}}
\nc{\norm}[1]{\|#1\|}
\nc{\abscut}[2]{\Abs{#1}_{\scriptscriptstyle#2}}
\nc{\vek}[1]{{\rm\bf #1}}
\nc{\integral}[2]{\int\limits_{#1}^{#2}}
\nc{\inv}[1]{\frac{1}{#1}}
\nc{\dd}[2]{{{\partial #1}\over{\partial #2}}}
\nc{\ddd}[2]{{{{\partial}^2 #1}\over{\partial {#2}^2}}}
\nc{\dddd}[3]{{{{\partial}^2 #1}\over
    {\partial #2 \partial #3}}}
\nc{\dder}[2]{{{d #1}\over{d #2}}}
\nc{\ddder}[2]{{{d^2 #1}\over{d {#2}^2}}}
\nc{\dddder}[3]{{d^2 #1}\over
    {d #2 d #3}}
\nc{\dx}[1]{d\,^{#1}x}
\nc{\dy}[1]{d\,^{#1}y}
\nc{\dz}[1]{d\,^{#1}z}
\nc{\dl}[1]{\frac{d\,^{#1}l}{(2\pi)^{#1}}}
\nc{\dk}[1]{\frac{d\,^{#1}k}{(2\pi)^{#1}}}
\nc{\dq}[1]{\frac{d\,^{#1}q}{(2\pi)^{#1}}}

\nc{\bfT}{{\bf T }}
\def\GeV{{\rm\ GeV}}

\nc{\cA}{{\cal A}}
\nc{\cB}{{\cal B}}
\nc{\cD}{{\cal D}}
\nc{\cE}{{\cal E}}
\nc{\cG}{{\cal G}}
\nc{\cH}{{\cal H}}
\nc{\cL}{{\cal L}}
\nc{\cO}{{\cal O}}
\nc{\cT}{{\cal T}}
\nc{\cN}{{\cal N}}
\nc{\rvac}[1]{|{\cal O}#1\rangle}
\nc{\lvac}[1]{\langle{\cal O}#1|}
\nc{\rvacb}[1]{|{\cal O}_\beta #1\rangle}
\nc{\lvacb}[1]{\langle{\cal O}_\beta #1 |}
\nc{\bb}{\bar{\beta}}
\nc{\bt}{\tilde{\beta}}
\nc{\ctH}{\tilde{\cal H}}
\nc{\chH}{\hat{\cal H}}
\nc{\1}{\aa}
\nc{\2}{\"{a}}
\nc{\3}{\"{o}}
\nc{\4}{\AA}
\nc{\5}{\"{A}}
\nc{\6}{\"{O}}
\nc{\al}{\alpha}
\nc{\g}{\gamma}
\nc{\Del}{\Delta}
\nc{\e}{\textrm{e}}
\nc{\eps}{\epsilon}
\nc{\lam}{\lambda}
\nc{\Om}{\Omega}
\nc{\ve}{\varepsilon}
\nc{\mn}{{\mu\nu}}
\nc{\vp}{\varphi}

\nc{\advp}[3]{{\it  Adv.\ in\ Phys.\ }{{\bf #1} {(#2)} {#3}}}
\nc{\annp}[3]{{\it  Ann.\ Phys.\ (N.Y.)\ }{{\bf #1} {(#2)} {#3}}}
\nc{\apl}[3]{{\it  Appl. Phys. Lett. }{{\bf #1} {(#2)} {#3}}}
\nc{\apj}[3]{{\it  Ap.\ J.\ }{{\bf #1} {(#2)} {#3}}}
\nc{\apjl}[3]{{\it  Ap.\ J.\ Lett.\ }{{\bf #1} {(#2)} {#3}}}
\nc{\app}[3]{{\it Astropart.\ Phys.\ }{{\bf #1} {(#2)} {#3}}}
\nc{\cmp}[3]{{\it  Comm.\ Math.\ Phys.\ }{{ \bf #1} {(#2)} {#3}}}
\nc{\cqg}[3]{{\it  Class.\ Quant.\ Grav.\ }{{\bf #1} {(#2)} {#3}}}
\nc{\epl}[3]{{\it  Europhys.\ Lett.\ }{{\bf #1} {(#2)} {#3}}}
\nc{\ijmp}[3]{{\it Int.\ J.\ Mod.\ Phys.\ }{{\bf #1} {(#2)} {#3}}}
\nc{\ijtp}[3]{{\it Int.\ J.\ Theor.\ Phys.\ }{{\bf #1} {(#2)} {#3}}}
\nc{\jmp}[3]{{\it  J.\ Math.\ Phys.\ }{{ \bf #1} {(#2)} {#3}}}
\nc{\jpa}[3]{{\it  J.\ Phys.\ A\ }{{\bf #1} {(#2)} {#3}}}
\nc{\jpc}[3]{{\it  J.\ Phys.\ C\ }{{\bf #1} {(#2)} {#3}}}
\nc{\jap}[3]{{\it J.\ Appl.\ Phys.\ }{{\bf #1} {(#2)} {#3}}}
\nc{\jpsj}[3]{{\it J.\ Phys.\ Soc.\ Japan\ }{{\bf #1} {(#2)} {#3}}}
\nc{\lmp}[3]{{\it Lett.\ Math.\ Phys.\ }{{\bf #1} {(#2)} {#3}}}
\nc{\mpl}[3]{{\it  Mod.\ Phys.\ Lett.\ }{{\bf #1} {(#2)} {#3}}}
\nc{\ncim}[3]{{\it  Nuov.\ Cim.\ }{{\bf #1} {(#2)} {#3}}}
\nc{\np}[3]{{\it  Nucl.\ Phys.\ }{{\bf #1} {(#2)} {#3}}}
\nc{\pr}[3]{{\it Phys.\ Rev.\ }{{\bf #1} {(#2)} {#3}}}
\nc{\pra}[3]{{\it  Phys.\ Rev.\ A\ }{{\bf #1} {(#2)} {#3}}}
\nc{\prb}[3]{{\it  Phys.\ Rev.\ B\ }{{{\bf #1} {(#2)} {#3}}}}
\nc{\prc}[3]{{\it  Phys.\ Rev.\ C\ }{{\bf #1} {(#2)} {#3}}}
\nc{\prd}[3]{{\it  Phys.\ Rev.\ D\ }{{\bf #1} {(#2)} {#3}}}
\nc{\prl}[3]{{\it Phys\ Rev.\ Lett.\ }{{\bf #1} {(#2)} {#3}}}
\nc{\pl}[3]{{\it  Phys.\ Lett.\ }{{\bf #1} {(#2)} {#3}}}
\nc{\prep}[3]{{\it Phys\. Rep.\ }{{\bf #1} {(#2)} {#3}}}
\nc{\prsl}[3]{{\it Proc.\ R.\ Soc.\ London\ }{{\bf #1} {(#2)} {#3}}}
\nc{\ptp}[3]{{\it  Prog.\ Theor.\ Phys.\ }{{\bf #1} {(#2)} {#3}}}
\nc{\ptps}[3]{{\it  Prog\ Theor.\ Phys.\ suppl.\ }{{\bf #1} {(#2)} {#3}}}
\nc{\physa}[3]{{\it  Physica\ A\ }{{\bf #1} {(#2)} {#3}}}
\nc{\physb}[3]{{\it  Physica\ B\ }{{\bf #1} {(#2)} {#3}}}
\nc{\phys}[3]{{\it Physica\ }{{\bf #1} {(#2)} {#3}}}
\nc{\rmp}[3]{{\it  Rev.\ Mod.\ Phys.\ }{{\bf #1} {(#2)} {#3}}}
\nc{\rpp}[3]{{\it Rep.\ Prog.\ Phys.\ }{{\bf #1} {(#2)} {#3}}}
\nc{\sjnp}[3]{{\it Sov.\ J.\ Nucl.\ Phys.\ }{{\bf #1} {(#2)} {#3}}}
\nc{\spjetp}[3]{{\it Sov.\ Phys.\ JETP\ }{{\bf #1} {(#2)} {#3}}}
\nc{\yf}[3]{{\it Yad.\ Fiz.\ }{{\bf #1} {(#2)} {#3}}}
\nc{\zetp}[3]{{\it Zh.\ Eksp.\ Teor.\ Fiz.\  }{{\bf #1}  {(#2)} {#3}}}
\nc{\zp}[3]{{\it Z.\ Phys.\ }{{\bf #1} {(#2)} {#3}}}
\nc{\ibid}[3]{{\sl ibid.\ }{{\bf #1} {#2} {#3}}}

\nc{\rf}[1]{(\ref{#1})}
\nc{\nn}{\nonumber \\*}
\nc{\bfB}{\bf{B}}
\nc{\bfv}{\bf{v}}
\nc{\bfx}{\bf{x}}
\nc{\bfy}{\bf{y}}
\nc{\vx}{\vec{x}}
\nc{\vy}{\vec{y}}
\nc{\oB}{\overline{B}}
\nc{\oI}{\overline{I}}
\nc{\oR}{\overline{R}}
\nc{\rar}{\rightarrow}
\nc{\ti}{\times}
\nc{\slsh}{\hskip-5pt/}
\nc{\sm}{Standard~Model~}
\nc{\MP}{M_{\rm Pl}}
\nc{\tp}{t_{\rm Pl}}

\nc{\pmin}{p_{\rm min}}
\nc{\pmax}{p_{\rm max}}
\nc{\fo}{f_0}
\nc{\foi}{f_{0,i}\,}
\nc{\fop}{f_0^P}
\nc{\fou}{f_0^U}

\nc{\eff}{{\rm eff}}
\nc{\MT}{M_{\rm T}}
\nc{\ML}{M_{\rm L}}
\nc{\kk}{\vek{k}}
\nc{\pp}{{\rm p}}
\nc{\pt}{\partial_t}
\nc{\half}{{1\over 2}}
\nc{\w}{\omega}
\nc{\uhat}{\hat{U}_\w}

\nc{\etal}{\mbox{\it et al.}}
\nc{\ie}{{\it i.e. }}
\nc{\eg}{{\it e.g. }}
\nc{\trh}{T_{\rm RH}}
\nc{\vo}{\varphi_\omega}
\nc{\G}{\tilde{G}}
\nc{\ansatz}{{\it Ansatz }}
\nc{\cm}{{\rm cm}}

\begin{document}

\title{{\hfill {{\small  TURKU-FL-P40-02
        }}\vskip 1truecm}
{\LARGE Limits on Q-ball size due to gravity}
\vspace{-.2cm}}

\author{{\sc Tuomas Multam\"aki\footnote{email: tuomul@utu.fi}}\\ 
and\\
{\sc Iiro Vilja\footnote{email: vilja@utu.fi}}\\ 
\\{\sl\small Department of Physics, University of Turku, FIN-20014, FINLAND}}

\date{May 28, 2002}

\maketitle
\thispagestyle{empty} 

\abstract{
Solitonic scalar field configurations are studied in a theory coupled
to gravity. It is found that non-topological solitons, Q-balls,
are present in the theory. Properties of gravitationally self coupled 
Q-balls are studied by 
analytical and numerical means. Analytical arguments show that,
unlike in the typical flat space scenario,
the size of Q-balls is ultimately limited by gravitational effects.
Even though the largest Q-balls are very dense, their radii are still much 
larger than the corresponding Schwarzschild radii.
Gravity can also act as a stabilising mechanism for otherwise energetically
unstable Q-balls.
}

\setlength{\captionmargin}{30pt}

\newpage
\setcounter{page}{1} 

\section{Introduction}

Various field theories can support stable non-topological solitons
\cite{leepang}, Q-balls \cite{coleman}, which 
are coherent scalar
condensates that carry a conserved charge, typically a $U(1)$ charge.
Due to charge conservation, the Q-ball configuration is the ground state
in the sector of fixed charge. 
Q-balls may have physical importance because the
supersymmetric extensions of the Standard Model have
scalar potentials that are suitable for Q-balls to exist in the theory. 
In particular, lepton or baryon number carrying Q-balls
are present in the Minimal Supersymmetric Standard Model (MSSM) due 
to the existence of flat directions in the scalar sector
of the theory \cite{kusenko405,misha418,kari538,kari425}.

Q-balls can be cosmologically significant in various ways. 
Stable (or long living) Q-balls are natural candidates
for dark matter \cite{misha418} and their decay offers 
a way to understand the baryon to dark matter ratio
\cite{kari538} as well as the baryon asymmetry of the universe
\cite{kari538,kari425}. Q-balls can also protect the baryons
from electroweak sphalerons \cite{kari425} and may
be an important factor in considering the stability of 
neutron stars \cite{wreck}.

For Q-balls to be cosmologically significant, one needs
to have a mechanism that creates them in the early 
stages of the evolution of the universe. Q-balls can be efficiently created
in the early universe from an Affleck-Dine (AD) condensate 
\cite{misha418,kari538}. This process has been
studied by numerical simulations \cite{kasuya1}-\cite{3dsims} 
where both the gauge- and gravity-mediated SUSY breaking scenarios
were considered. In both cases, Q-balls with various charges
have been seen to form from the condensate.
Also collisions of Q-balls have been considered in various 
potentials \cite{kasuya2,colli}.

A large Q-ball is a very dense object and hence
its coupling to the gravitational field may be remarkable. In the present
paper we study these aspects of Q-ball solutions \ie we are interested in
the gravitational coupling of scalar configurations. Some such considerations
have also been studied with reference to Q-stars \cite{lynn}. 

The present paper is organised as follows: In Chapter 2 we briefly review the
properties of Q-balls in flat space and study analytically the
gravitationally coupled system. In Chapter 3 special attention is 
given to gravitationally coupled thin-walled Q-ball configurations. 
Numerical results are presented in Chapter 4. The paper is 
concluded in Chapter 5.

\section{Theoretical considerations}
\subsection{Preliminaries: Q-balls in flat space}

Consider a field theory with a U(1) symmetric scalar potential,
$V(|\phi|)$, with a global minimum at $\phi=0$. The complex scalar field
$\phi$ carries a unit quantum number with respect to the $U(1)$-symmetry.
The charge and energy of a field configuration $\phi$ 
in D dimensions are \cite{leepang}
\be{charge}
Q={1\over i}\int (\phi^*\dot\phi-\phi\dot\phi^*)d^Dx
\ee
and
\be{energy}
E=\int [|\dot{\phi}|^2+|\nabla\phi|^2+V(\phi^*\phi)]d^Dx.
\ee
The Q-ball solution is the minimum energy configuration 
in the sector of fixed charge. A Q-ball is stable
against radiative decays into $\phi$-scalars if condition
\be{stabcond}
E<mQ,
\ee
where $m$ is the mass of the $\phi$-scalar, holds. It is then
energetically favourable to store charge in a Q-ball rather
than in the form of free scalars.

Finding the minimum energy is straightforward and 
the Q-ball solution can be shown to be of the form
\cite{leepang,coleman}
\be{ansatz}
\phi(x,t)=e^{i\omega t}\varphi(r),
\ee
where $\varphi(x)$ is now time independent and real, $\omega$ is 
the Q-ball frequency (for most cases $|\omega|\in[0,m]$).
The charge of a Q-ball with spherical symmetry in D-dimensions is
given by
\be{sphercharge}
Q=2\omega\int \phi(r)^2 d^Dr
\ee
and the equation of motion at a fixed $\omega$ is 
\be{eom}
{d^2\phi\over dr^2}+{D-1\over r}{d\phi\over dr}=\phi{\partial
U(\phi^2)\over\partial\phi^2}-\omega^2\phi.
\ee 
To find the Q-ball solution we must solve (\ref{eom}) with
the boundary conditions $\phi'(0)=0,\ \phi(\infty)=0$.

The solution of Eq. (\ref{eom}) depends crucially on the shape of 
the potential. Flat potentials, which are essentially constant over 
wide range of field $\varphi$ give rise to the relation $E \propto Q^{3/4}$ 
\cite{dvali417}, whereas in a wide class of more general potentials
the energy-charge relations is  linear, $E \propto Q$ \cite{leepang,coleman}. 
However, any Q-ball solution in a potential which is bounded from below,
asymptotically increases faster than $\varphi^2$ and has a global minimum
at the origin, approaches the thin-wall limit when $\omega$ becomes small 
enough. At the thin-wall limit any potential then leads to the linear relation 
between energy and charge.

\subsection{Field equations for gravitationally coupled Q-ball}

Supposing that the space time has a maximally symmetric subspace $S^2$, \ie
the configuration is spherically symmetric, the metric tensor can be written 
in the form
\be{metric}
g_{\mu\nu} = {\rm diag}(B,-A,-r^2,-r^2 \sin^2\theta),
\ee
where $A$ and $B$ are positive functions of time and radial coordinate \cite{wein}.
A complex scalar field $\phi$ coupled to gravity is then described by the 
action
\be{action}
S[\phi, A, B] = -\frac 1{16\pi G}\int d^4x \sqrt{AB}\,{\cal R} + 
\int d^4x \sqrt{AB}\left(\partial_\mu\phi^\dagger\partial^\mu\phi - 
V(\phi^\dagger\phi)\right ),
\ee
where ${\cal R}$ is the Ricci curvature scalar, $V(\phi^\dagger\phi)$ is the
scalar potential and $G=1/M_{Pl}^2$. By varying the action one obtains 
the well known Einstein equations together with the equation of motion 
for the scalar field. They read for the spherically symmetric system in 
the contracted form as
\bea{ceq}
-{B''\over 2 A} + {B'\over 4 A}\left ( {A'\over A} + {B'\over A} \right ) 
- {B'\over A r} + {\ddot A\over 2 A} - 
{\dot A\over 4 A}\left ( {\dot A\over A} + {\dot B\over A} \right ) = 
- 4\pi G (\rho +3 P)B,\nonumber\\
{B''\over 2 B} - {B'\over 4 B}\left ( {A'\over A} + {B'\over A} \right ) 
- {A'\over A r} - {\ddot A\over 2 B} + 
{\dot A\over 4 B}\left ( {\dot A\over A} + {\dot B\over A} \right ) =
- 4\pi G (\rho - P)A,\nonumber\\
 -1 + \frac 1A - {r\over 2 A}
\left({A'\over A} - {B'\over B}\right )= -4\pi G (\rho -P)r^2,\\
- {\dot A\over rA} = -4\pi G\left(\dot\phi^\dagger\phi' + \phi'^\dagger
\dot\phi\right ),\nonumber\\
\phi'' + \phi' \frac 1{r^2} \sqrt{\frac AB}{\partial\over \partial r}\left (
r^2\sqrt{\frac BA}\right ) - A{\partial V\over \partial \phi^\dagger} =
\frac AB \left ( \ddot\phi + \dot\phi \sqrt{\frac BA}{\partial\over \partial t}
\sqrt{\frac AB}\right ),\nonumber
\eea
where the energy density $\rho$ and pressure $P$ are given by
\bea{rhojaP}
\rho = \frac 1B |\dot\phi|^2 + \frac 1A |\phi'|^2 + V,\nonumber\\
P=  \frac 1B |\dot\phi|^2 + \frac 1A |\phi'|^2 - V.
\eea
The total energy $E_{\rm tot}$ of the Q-ball, including the gravitational 
energy, is still given by the formula
\be{Etot}
E_{\rm tot} = \int d^3x\rho,
\ee
whereas the energy of the scalar configuration without the gravitational
contribution, $E_{\rm s}$, can be defined as
\be{Es}
E_{\rm s} = \int d^3x\, \sqrt {AB}\rho.
\ee
Of course, $E_{\rm tot}$ is the true energy of the configuration, but $E_{\rm s}$
is comparable to the corresponding flat space configuration, as will be shown. 
In the presence of gravity also the formula for the charge of the Q-ball 
is modified and is now given by
\be{q1}
Q= -i\omega\int d^3x\sqrt {AB}\left ( \phi^\dagger\dot\phi - \dot\phi^\dagger\phi\right ),
\ee
which differs from the flat space formula by the measure factor $\sqrt {AB}$.

From the set of equations (\ref{ceq}) one can see that the {\it Ansatz} (\ref{ansatz}) 
for the scalar field is still applicable. The right hand side of the equation 
for $R_{tr}$ then vanishes, implying that $A$ is independent on time.
Then, as usual, we can without a loss of generality assume that also $B$ 
is time independent. The new set of equations now reads as
\bea{eqs}
-{B''\over 2 A} + {B'\over 4 A}\left ( {A'\over A} + {B'\over A} \right ) 
- {B'\over A r} = - 4\pi G (\rho +3 P)B,\nonumber\\
{B''\over 2 B} - {B'\over 4 B}\left ( {A'\over A} + {B'\over A} \right ) 
- {A'\over A r}= - 4\pi G (\rho - P)A,\\
-1 + \frac 1A - {r\over 2 A}
\left({A'\over A} - {B'\over B}\right )= -4\pi G (\rho -P)r^2\nonumber
\eea
and
\be{phieq}
\varphi'' + \varphi' \frac 1{r^2} \sqrt{\frac AB}{\partial\over \partial r}\left (
r^2\sqrt{\frac BA}\right ) - AV'_\omega(B,\varphi) = 0\nonumber.
\ee
The $\omega$-dependent potential $V_\omega$ is defined as
\be{vo}
V_\omega(B,\varphi)= \frac 12 V(\varphi^2) - \frac 12  {\omega^2\over B} \varphi^2
\ee
and $V'_\w$ denotes a derivate of the potential with respect the field $\varphi$.
The charge of a Q-ball is now given by
\be{q2}
Q=8\pi\omega\int_0^\infty dr\, r^2\sqrt {A(r)B(r)}\, \varphi^2.
\ee

Like in the flat case, a necessary but not a sufficient condition for the
existence of Q-balls is that the $\omega$-dependent potential
has a non-zero minimum $\varphi_+$ with $V_\omega(B,\varphi_+)<V_\omega(B,0)$
\footnote{From now on we suppose, that $V_\omega(0)=0$. We can also choose, 
without a loss of generality, the central values of $A$ and $B$ to be $A(0)=B(0)=1$. 
In addition, for finite energy field configurations $A(\infty ) = 1$.}.
As usual, this restricts the values of $\omega$, but it also constrains the 
function $B$, because $V_\omega$ depends explicitly on it. The allowed values 
for $\omega$ are hence restricted to be larger than a critical, non-negative value 
of $\omega_c$, {\it i.e.} $\omega > \omega_c$. The critical value, $\w_c$, is
determined by the set of equations
\bea{oc}
\begin{array}{ccc}
V_{\omega_c}(1,\varphi_c) & = & 0,\\
V'_{\omega_c}(1,\varphi_c) & = & 0.
\end{array}
\eea

The equations for $A$ and $B$ (\ref{eqs}) can be easily solved
in terms of $\rho$ and $P$:
\be{eqforA}
\left ({r\over A}\right )' = 1 - 8\pi G r^2\rho,
\ee
\be{eqforB}
{B'\over B} + {A'\over A} = 8\pi G A r(\rho+P).
\ee
These equations can not, however, be integrated straightforwardly because both
$\rho$ and $P$ are dependent on $A$, $B$ and the field $\varphi$. Therefore our 
aim in the following sections is twofold:
In the next section we study the equations and their solutions by utilising
analytical approximations in order to gain an understanding of their fundamental 
properties. To verify these results, Eqs. (\ref{eqs}) are studied numerically 
in Section \ref{NR}.

\section{The thin-wall limit and gravity}

If the $\omega$-dependent potential $V_\omega$ is such that it fulfils the
conventional thin-wall requirements, \ie the height of the barrier is much
larger than the difference between minima, the field forms a bubble of radius $R$. 
Inside the bubble the value of the field is essentially constant and 
outside virtually zero with only a 
thin transition region separating them. The metric components $A$ and $B$ are, however,
not constants, but monotonic functions inside the bubble. Thus, one has to
determine their behaviour directly from Eqs. (\ref{eqforA}) and (\ref{eqforB}).
To determine the properties of such a configuration, it is useful to consider
the effective action
\be{ea}
S_\omega = 4\pi \int_0^\infty dr\,r^2\sqrt {A(r)B(r)}\, \left (
\frac 1{2A(r)}\varphi'(r)^2 + V_\omega(B(r),\varphi(r))\right ),
\ee
from which the equation of motion for $\varphi$, Eq. (\ref{phieq}), emerges.
Applying the usual thin-wall approach to the action (\ref{ea}), leads to some complications
due to the presence of gravity. From Eq. (\ref{oc}) we see that two separate 
sub-cases need to considered, depending on whether the potential $V$ is monotonic 
or it has a non-zero local minimum. Defining $\varphi_+$ by 
\be{maxeq}
V'_{\omega_c}(1,\varphi_+)=0,\ V''_{\omega_c}(1,\varphi_+) < 0,
\ee
\ie $\varphi_+$ determines the location of barrier maximum, 
we can now make difference between type I and type II thin-wall Q-balls. 
A thin-wall solution is classified as type I if 
$V_{\omega_c}(1,\varphi_+)\simeq   
\omega_c^2 \varphi_c^2$ and type II if $V_{\omega_c}(1,\varphi_+) \gg
\omega_c^2 \varphi_c^2$, where $\varphi_c$ is determined by Eqs. (\ref{oc}).
Type I thin-wall hence corresponds to a monotonically increasing
potential $V$ whereas type II thin-wall solutions appear when the potential
has a non-zero local minimum such that it is nearly degenerate with minimum at
$\varphi=0$.

In both cases we assume that inside the bubble $\varphi$ is essentially
constant. Its value $\varphi_\omega$ is determined
by equation
\be{phio}
V'_\omega(B(R),\varphi_\omega)=0.
\ee
Note that even though $B(r)$ is varying inside the bubble, the change
is relatively small which justifies our approximation, $B(r)\approx B(R)$.

The value of $S_\omega$ inside the bubble is hence given by
\be{inside}
S_{\omega,in} = 4\pi \int_0^R dr\, r^2\sqrt {AB}\,  V_\omega(B(r),\varphi(r)).
\ee 
Note, that now $\omega > \omega_c$ and thus $V_\omega(B(r),\varphi(r))<0$
and $E_{\omega, in}$ is negative.
Note also that due to Eq. (\ref{phio}), $\varphi_\omega$ depends 
on the bubble radius $R$.
This negative contribution to $E_\omega$ should be balanced by the
surface term. If $A(R)$ and $B(R)$ differ only little from unity, the wall 
contribution can be approximated by 
\be{wall}
S_{\omega, wall} \simeq 4\pi\sigma R^2,
\ee
where the surface tension is determined by varying the action with the
degenerated potential $V_\omega$. In practice this means that, as a first
approximation, we set $\omega=\omega_c,\ B=A=1$, and we finally end up with the 
usual formula for the surface tension:
\be{sigma}
\sigma \simeq \int _0^{\varphi_c} \sqrt {2V_{\omega_c}(1, \varphi)}\ d\varphi.
\ee
Therefore the effective action, reads out as
\be{twa}
S_\omega = S_{\omega,in} + S_{\omega, wall}.
\ee
The difference between Eq. (\ref{twa}) and the usual thin-wall case lies in the
explicit dependence of Eq. (\ref{twa}) on the metric components $A$ and $B$.
The next step is then to solve $A$ and $B$ from Eqs. (\ref{eqforA}) and
(\ref{eqforB}). Note, that in this approximation the energy density
\be{rho2}
\rho = \frac {\omega^2}B |\varphi|^2 + \frac 1A |\varphi'|^2 + V(\varphi^2)
\ee
is virtually constant inside the bubble, where one can also neglect the 
derivative term. 

From this point on the two bubble types corresponding to the different
types of thin-wall potentials differ from each other
and are discussed separately.

\subsection{Type I thin-wall solution}

For the type I solution the terms $\frac{\omega^2}B |\varphi|^2$ 
and $V(\varphi^2)$ are essentially equal. The relative variation of $\rho$
is also small and hence it is essentially constant. 
We can thus integrate $A$ out from Eq. (\ref{eqforA}) and obtain
\be{A}
A(r)={1\over 1-{8 \pi\over 3} G \rho r^2}={1\over 1-{16 \pi\over 3} G 
V(\varphi_\omega^2) r^2}.
\ee
Next we note that within the same approximation $\rho + P = \frac {2 \omega^2}B |\varphi|^2  
+ \frac 2A |\varphi'|^2 \simeq \frac {2 \omega^2}B |\varphi|^2$
and Eq. (\ref{eqforB}) can be rewritten as
\be{eqforB2}
B' + \frac {A'}A B = 16\pi G \omega^2 \varphi^2 A r,
\ee
which can be solved in the the leading order of $G$,
\be{B1}
B(r) = 1 + {8\pi\over 3}G V(\varphi_\omega^2) r^2.
\ee

We are now able to write down the equations determining the gravitationally coupled thin-wall
solution for the Q-balls. We use the solution given above for $B(r)$, solve the relation 
between $R$ and $\varphi_\omega$ using (\ref{phio}) and calculate $R$ from 
${d E_\omega\over dR} = 0$. So, in addition to Eqs. (\ref{A}) and (\ref{B1}),
we have equations
\be{set}
V'_\omega(B(R),\varphi_\omega)=0,\, 4\pi R^2 \sqrt {A(R)B(R)}\,V_\omega(B(R),\varphi_\omega) + 8\pi\sigma R=0,
\ee
which determine the value of $\omega$ as well as $R$ and the corresponding value
of the field, $\varphi_\omega$.
Using these values for $\omega$, $R$ and $\varphi_\omega$
we may also calculate the corresponding 
charge $Q$ from Eq. (\ref {q2}). If $A$ and $B$ do not differ too much from 
unity we simply obtain
\be{q3}
Q={8\pi\over 3} \omega \varphi_\omega^2 R^3,
\ee
where we have neglected the contribution from the wall.

The new features of the gravitationally coupled Q-balls arise due to the non-trivial
dependence of the energy inside the bubble to the bubble radius $R$. 
Suppose that $\omega$ is close enough $\omega_c$. The scale factor $B(R)$ is 
a monotonically increasing function, and it may happen that at some point the 
effective $\omega$-parameter $\omega(R)=\omega/\sqrt {B(R)}$ becomes too small
and the Eqs. (\ref{set}) cannot be satisfied.  
Because $R$ and $\varphi_\omega$ depend implicitly on $\omega$, we are now able to 
define another critical value of $\omega$, $\omega_G$, as the smallest value where a 
solution of Eqs. (\ref{set}) can be found. Thus when gravity effects are taken into
account, Q-balls seem to exist only if $\omega > \omega_G$ which, on the other hand,
implies maximal a radius $R_{max}$ and maximal charge $Q_{max} = {8\pi\over 3} \omega_G
\varphi_{\omega_G}^2R_{max}^3$. The only other possibility is, that the singular
point of $A(R)$ is hit first. That, however, would mean that such a configuration would
collapse and form a black hole and hence such Q-balls are not possible either.
Clearly, the critical values $\omega_{\rm min}$, $R_{max}$ and $Q_{max}$ depend explicitly
on the shape of the potential and to proceed further one has to define it more precisely.

To see explicitly how the Q-ball charge is constrained, we study a simple example potential.
The model potential, $V$, is constant, $m^4$, for $\varphi<\varphi_1$, 
expect near the origin where it tends 
to zero as $\varphi^2$. The exact behaviour near the origin is not important in
our analysis. For values $\varphi > \varphi_1$ the potential increases rapidly,
\eg $V \sim \varphi^n$ where the power $n$ is sufficiently large.

Near $\w_G$, the minimum of $V_\w$ is essentially $\varphi_1$.
The metric components are approximately given by
\bea{Iequs}
A(R) & = & 1+{16\pi\over 3}G V(\varphi_1^2)R^2\nonumber\\
B(R) & = & 1+{8\pi\over 3} G V(\varphi_1^2)R^2\nonumber
\eea
where we have dropped terms higher or equal to $\cO(G^2)$.
Within the same approximation, we get from Eq. (\ref{set})
an equation for $R$:
\be{rcube}
{2\pi G m^4\over 3} R^3(3m^4-\w^2\varphi_1^2)+{1\over 2}R(m^4-\w^2\varphi_1^2)+
2\sigma  = 0.
\ee
Only one of the roots of the cubic equation (\ref{rcube}) is physical
and is increasing with decreasing $\w$ until the critical point, $\w=\w_G$,
where the discriminant of the equation vanishes. The maximal value of
$R$ is
\be{Imaxr}
R_{max}=\Big({3\sigma\over 2\pi G m^4(3m^4-\w_G^2\varphi_1^2)}\Big)^{1/3}.
\ee
The corresponding charge and energy of the configuration are
\bea{Ieandq}
Q_{max} & \approx & 2({\varphi_1\over m})^2{1\over m^2G}\\
E_{max} & \approx & 2 {\varphi_1\over m} {1\over m G},
\eea
where we have assumed that $\w_G^2\varphi_1^2\approx m^4$ 
and used the approximation $\sigma \sim m^2 \varphi_1$.
The energy to charge ratio of the maximal Q-ball is $E_{max}/mQ_{max}=m/\vo$.
In order to see that our approximation is valid, we note that 
$Gm^4R_{max}^2\sim (\varphi_1/\MP)^{2/3}\ll 1$.


\subsection{Type II thin-wall solution}

For the type II solution the fine tuning $V_\omega(B,\phi)\simeq 0$ is not needed 
in order to reach the thin wall limit because a large barrier exists even 
without the $\omega^2\phi^2$ -term. Now, indeed,
$\omega^2\varphi_\omega^2 \gg V(\varphi_\omega^2)$ and we write an effective
energy density as $\rho = (1 + \kappa ) \omega^2 \varphi^2$, where
$\kappa$ is assumed to be a constant. Instead of equation (\ref{A}) we now write
\be{A2}
A(r)={1\over 1-{8 \pi\over 3} G \rho r^2}=
{1\over 1-{8 \pi\over 3} G (1+\kappa )\omega^2\varphi^2 r^2}
\ee
and from Eq. (\ref{eqforB2}) we obtain by integrating 
\be{B2}
B(r) = 1 + {8\pi\over 3}G (2-\kappa) \omega^2\varphi^2 r^2.
\ee
The difference between type I and II solutions
is that in the type I solution, $A$ increases near the origin twice as fast as
$B$ whereas for the type II solution the roles are reversed: $B$ increases roughly twice
as fast as $A$ whenever $\kappa \ll 1$. 

Due to exactly same reason as before, a minimal $\omega$ appears together with a
maximal bubble radius, charge and energy in this case as well. Generally, they can again be 
found solving Eqs. (\ref{set}), (\ref{A2}) and (\ref{B2}) and searching for the smallest possible
$\omega$. However, the procedure is strongly dependent on the explicit form of the 
potential $V$ as it was in the case of type I solutions.
 
Therefore, as an example, we again proceed by studying a simple potential
\be{pot1}
V(\varphi^2)= \lambda \varphi^2\left ( (\varphi - M)^2 + \epsilon M^2\right ),
\ee
where $M$ is a mass scale and $\epsilon \ll 1$ is a small positive parameter 
lifting the non-zero minimum up from $V=0$. Note, that the quadratic term 
near the origin reads as $\lambda(1+\epsilon )M^2 \varphi^2$, so that the mass 
of a single quantum is given by $m^2 \simeq \lambda M^2$. Calculating the 
surface energy, we find that $\sigma = \sqrt \lambda M^3/6$.
The set of equations (\ref{set}) reads now as:
\bea{eset}
\lambda (\vo -M)(2\vo - M)-{\omega^2\over B(R)}+\omega_c^2 & = & 0,\\
\sqrt{A(R)B(R)}\Big(\lambda \vo^2(\vo - M)^2 + (\omega_c^2-{\w^2\over B(R)}) \vo^2\Big)+{2\sigma\over R} & = & 0,
\eea
where $\omega_c^2 = \lambda \epsilon M^2$.  

Evidently $\w_G$ is close to $\w_c$ and $\vo$ is close to $M$, which allows
us to write $\vo=M$ and $\w=\w_c$ wherever no explicit difference
appears in the formulae. Expanding in the first power of $G$ we arrive at an 
equation for the bubble radius,
\be{IIrequ}
R^3-{3(\w^2/\w_c^2-1)\over 8\pi G\w_c^2M^2(2-\kappa)}R+{3\sigma\over 2\pi G \w_c^4M^4(2-\kappa)}=0.
\ee
Following the same arguments as in the type I case, we can solve for $R_{max}$ and $\w_G$:
\bea{IIrm}
R_{max} & = & \Big({\sqrt{\lambda}\over 4\pi(2-\kappa)}{\MP^2\over M\w_c^4}\Big)^{1/3}\\
\w_G^2 & = & \w_c^2\Big(1+2((2-\kappa)\pi {M^4\over \MP^2\w_c^2})^{1/3})\label{cd1}.
\eea
The approximations we made in the process of deriving the above equations
constrain the acceptable range of parameter values, $M\ll\sqrt{\MP\w_c}$. If this 
condition is not satisfied our results are not valid and the equations need to be 
solved numerically.

The maximal charge and energy of a Q-ball in this scenario are
\bea{IIqande}
Q_{max} & \approx & {2\sqrt{\lambda}\over 3(2-\kappa)} {M\MP^2\over \w_c^3}\\
E_{max} & \approx & {2\sqrt{\lambda}\over 3(2-\kappa)}{M\MP^2\over\w_c^2}.
\eea
The energy to charge ratio, $E_{max}/mQ_{max}=\w_c/m=\sqrt{\epsilon}/\ll 1$,
indicates that the maximal Q-ball is energetically stable.


\section{Numerical Results}\label{NR}

Next we turn to look at some numerical results obtained by computer simulations.
We concentrate ourselves to the two different potentials, one with a flat plateau
and one with a nearly degenerate minimum. 
The flat potential allows type I thin wall Q-ball solutions near the critical case. 
Such potentials are hence called type I potentials. The other type of 
potential is the one given by Eq. (\ref{pot1}), leading to the type II thin-walls
solutions and are therefore called type II potentials.
Note, that both of these potential types are represented, {\it e.g.}, in the MSSM,
where the potential arising from gauge mediated supersymmetry breaking is clearly
a type I potential whereas gravity mediated supersymmetry breaking typically
leads to a type II potential.

From the simulations we have learned, however, that the thin wall limit is
exceptionally difficult to study by simulation due to an extreme fine tuning 
required
from the initial value of the field inside the bubble. This fine tuning is
much more severe in the gravitationally coupled case, because here one is not attempting
to find just the field value, but also a corresponding value of $\omega$.
Therefore the simulations are concentrated to the thick-wall end of the parameter
space, \ie to larger $\omega$. At best we are able to to see some
hints of the beginning thin-wall behaviour. Also, in order to clearly demonstrate
the observed behaviour of the solutions, we use large mass 
scales in the potential, since otherwise the true thin-wall solution is needed.

\subsection{Type I potential}

\begin{figure}[ht]
\leavevmode
\centering
\vspace*{62mm}
\begin{picture}(0,0)(0,490)
\includegraphics{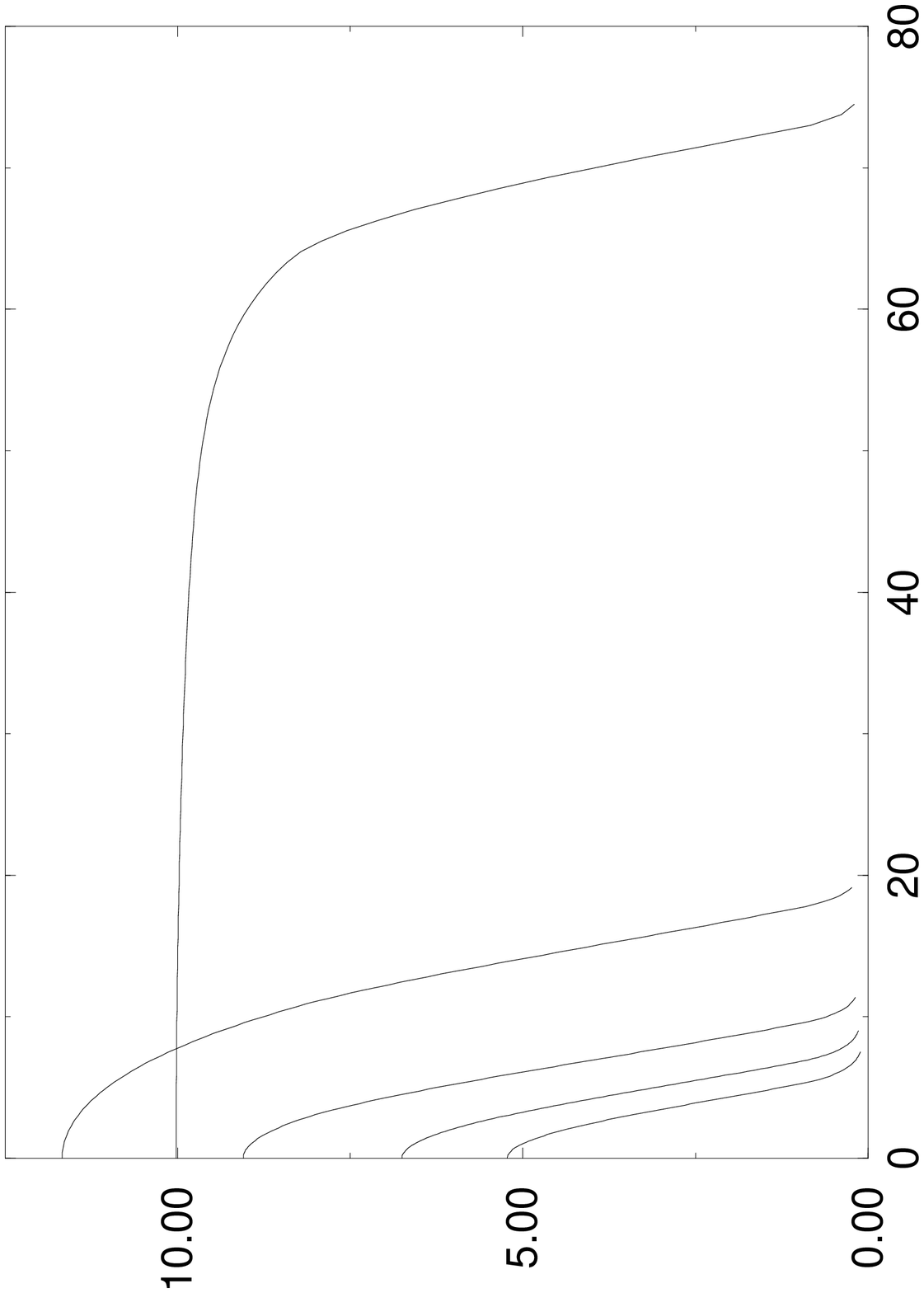}
\put(-110,308){$\scriptstyle r m$}
\put(-215,410){$\scriptstyle \varphi/m$}
  \put(-50,460){(a)}
\includegraphics{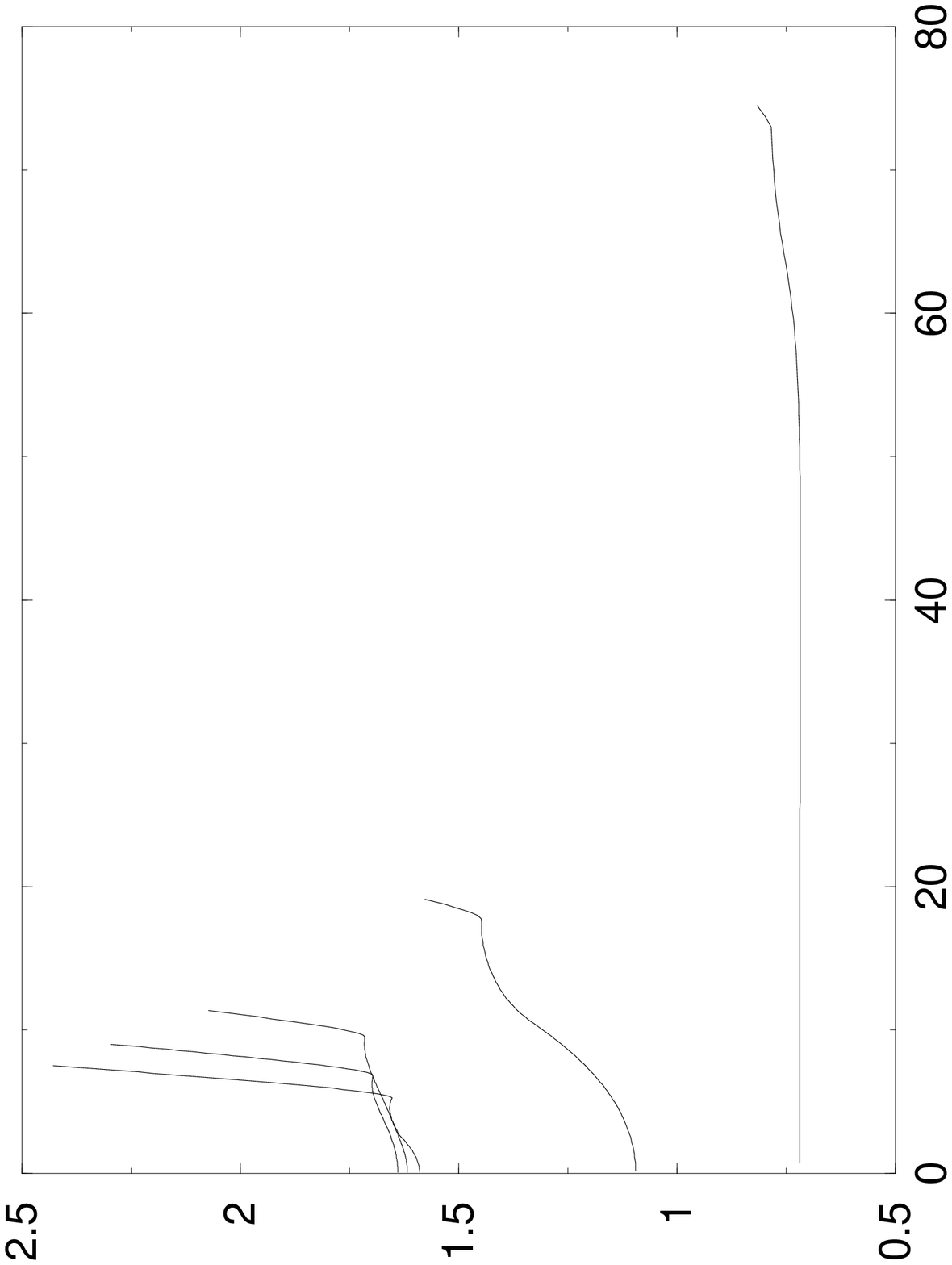}
\put(110,308){$\scriptstyle r m$}
\put(7,410){$\scriptstyle {B-1\over A-1}$}
 \put(170,460){(b)}
\end{picture}
\caption{(a) The field profile as function of radial coordinate. (b) The ratio of
metric components $(B-1)/(A-1)$. In the both figures curves represent values 
$\omega/m= 0.5$ (the shortest curve), $0.4,\ 0.3,\ 0.2$ and $0.142$ (the longest curve).
  \label{kuva1}
}
\end{figure}

\begin{figure}[ht]
\leavevmode
\centering
\vspace*{55mm}
\begin{picture}(0,0)(0,490)
\includegraphics{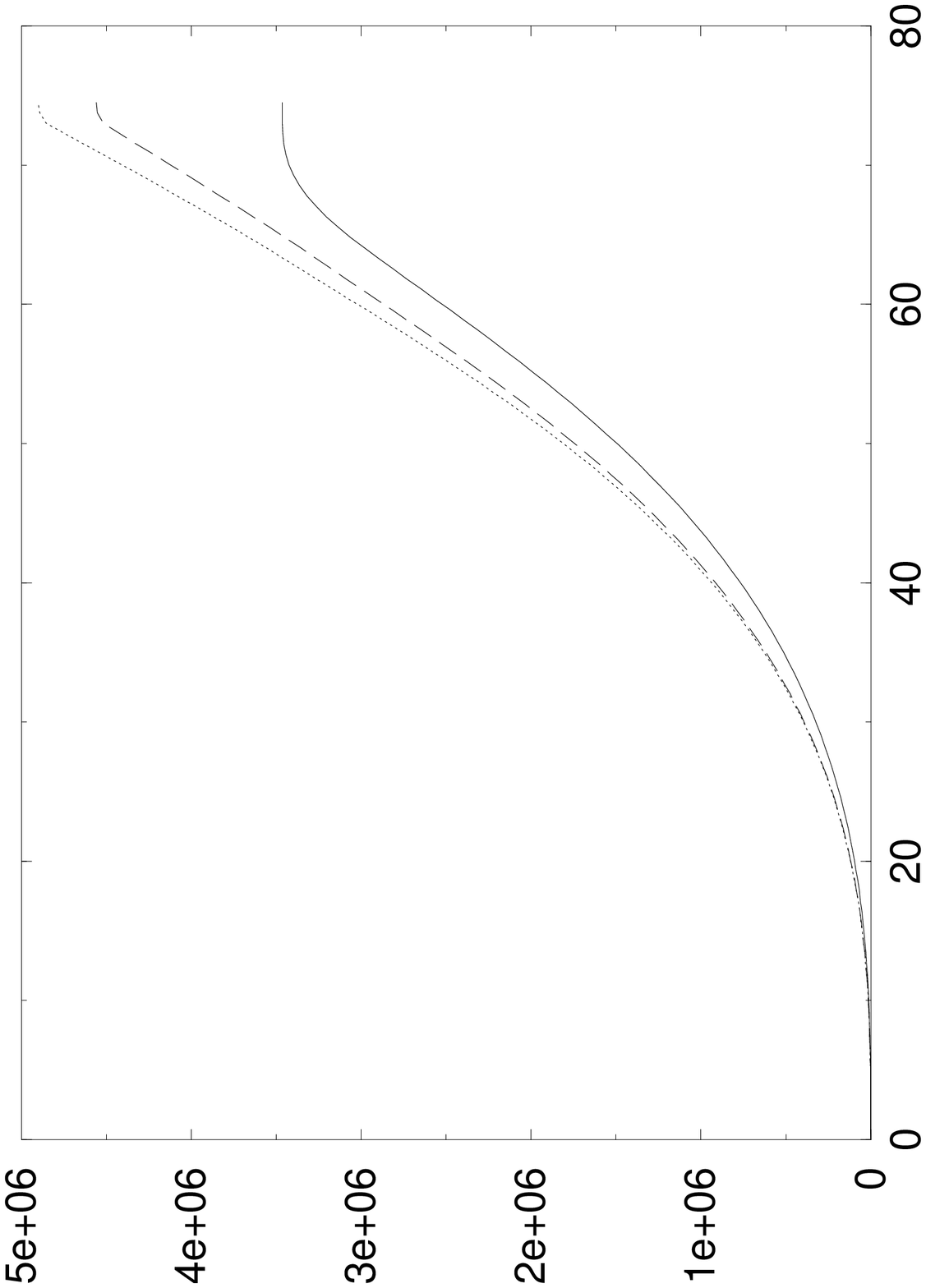}
\put(-10,325){$\scriptstyle r m$}
\put(-180,420){$\scriptstyle E/m$}
\end{picture}
\caption{The scaled energy of free quanta $0.1\times mQ/m$ (solid line), the total energy
$E_{\rm tot}/m$ (dashed line) 
and the scalar field energy $E_{\rm s}/m$ (dotted line) inside radius $r m$.}\label{kuva2}
\end{figure}

We begin with a potential
\be{numpotI}
V(\varphi^2) = \left \{ 
\begin{array}{lll}
m^2\varphi^2, & |\varphi | < \varphi_0\\
m^2\varphi^2_0, & \varphi_0 < |\varphi| < \varphi_1\\
m^2 \varphi^2_0 + \epsilon (\varphi^4 - \varphi^4_1), & \varphi_1 < \varphi
\end{array} \right.
\ee
where $m^2$ determines the curvature at the origin, $\epsilon$ is a small 
positive number and $0 < \varphi_0\ll\varphi_1$. To ensure that the minimum is
in the region $\varphi > \varphi_1$ and $\omega$ is larger that the critical
value, we have to require $m^2 > \omega^2 > 2\epsilon \varphi^2_1$ and
$4\epsilon (m^2 \varphi^2_0 - \epsilon \varphi^4_1) < \omega^4$, respectively.
This potential resembles the flat one discussed in the context of thin-wall 
configurations.
As an example we choose $m = \varphi_0 = 10^{-3}\MP,\ 
\varphi_1^4 = \frac 12 10^{-8}\MP^4$, and $\epsilon = 10^{-4}$. 
For these values, the critical 
$\omega$ equals $\omega_c = \sqrt 2 \times 10^{-1} m$ and the critical field value
is just $\varphi_1$.
The configuration is studied for several values of $\omega$ in order
to demonstrate the dependence of the solution on $\w$. Hence, we take 
$\omega/m = 0.5,\ 0.4,\ 0.3,\ 0.2$ and $0.142$. 
In Fig. \ref{kuva1}a we have plotted the field configuration of 
$\varphi$ and in Fig. \ref{kuva2}b the ratio $(B(r)-1)/(A(r)-1)$ 
for these values. Furthermore, in Fig. \ref{kuva2} we have plotted 
the corresponding energy of the free quanta $m Q$,
total energy $E_{\rm tot}$ and scalar field energy $E_{\rm s}$
as a function of $r$ for the largest Q-ball corresponding $\omega/m=0.142$.
Note that in Fig. \ref{kuva2}, the energy of the free quanta is
scaled by a factor of ten, \ie $E_{\rm tot},\ E_{\rm s}\ll mQ$.

Evidently, for large $\omega$, the profile of $\varphi$ is thick-walled. However,
from the figure it is clear that when $\omega$ decreases the configuration tends towards
the thin-wall case. This is also reflected by the fact that the ratio $(B(r)-1)/(A(r)-1)$
decreases along with decreasing $\omega$. Indeed, according to our analysis of
the type I thin-wall configurations, inside the bubble the ratio should approach 
$\frac 12$. 
As can bee seen from Fig. \ref{kuva2} the energy difference between 
$E_{\rm tot}$ and $E_{\rm s}$
is already for $\omega/m=0.142$ about 7\% and evidently it is increasing with 
increasing Q-ball size. It should be noted, that the critical $\omega$ which
is now $\omega_c/m \simeq 0.1414$, is already quite close to the
smallest value used in plotting the figure. In the large Q-ball limit, however, 
the configuration is extremely sensitive to the value of $\omega$.

\subsection{Type II potential}

\begin{figure}[ht]
\leavevmode
\centering
\vspace*{62mm}
\begin{picture}(0,0)(0,490)
\includegraphics{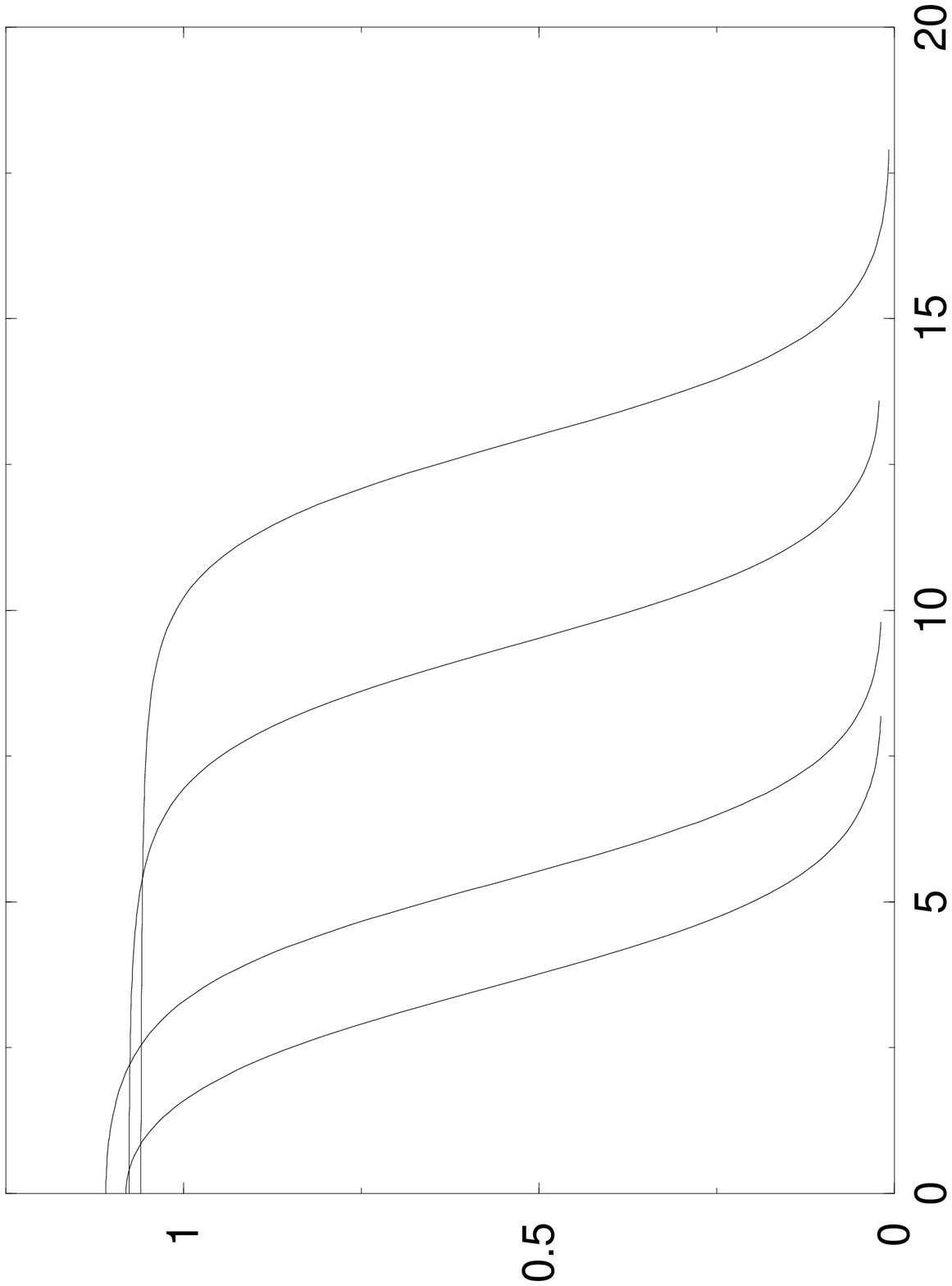}
\put(-110,308){$\scriptstyle r m$}
\put(-215,410){$\scriptstyle \varphi/m$}
  \put(-50,460){(a)}
\includegraphics{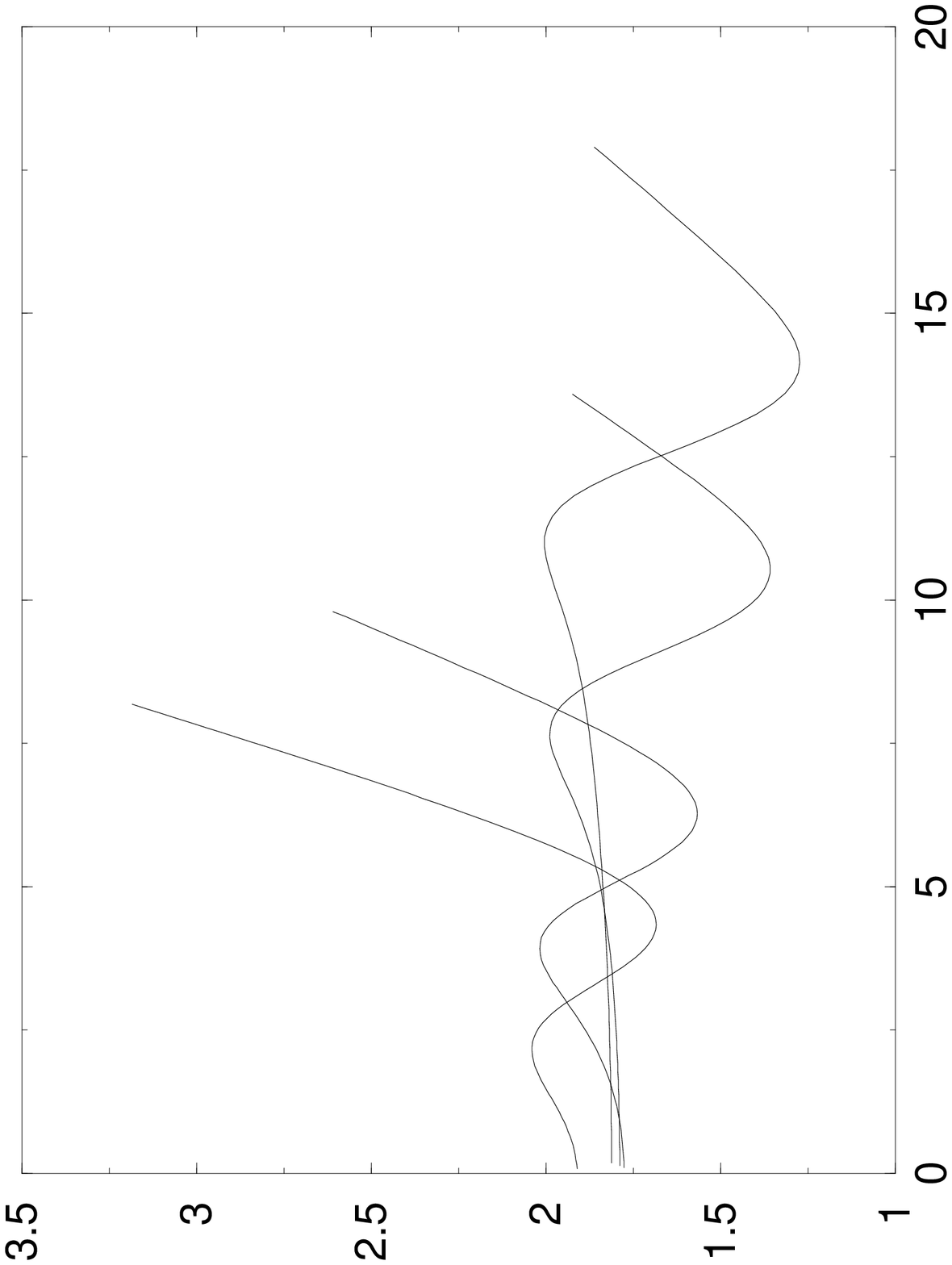}
\put(110,308){$\scriptstyle r m$}
\put(7,410){$\scriptstyle {B-1\over A-1}$}
 \put(170,460){(b)}
\end{picture}
\caption{(a) The field profile as function of radial coordinate. (b) The ratio of
metric components $(B-1)/(A-1)$. In the both figures curves represent values 
$\omega/m = 0.5,\ 0.4,\ 0.3$ and $0.261$.
  \label{kuva3}
}
\end{figure}

\begin{figure}[ht]
\leavevmode
\centering
\vspace*{55mm}
\begin{picture}(0,0)(0,490)
\includegraphics{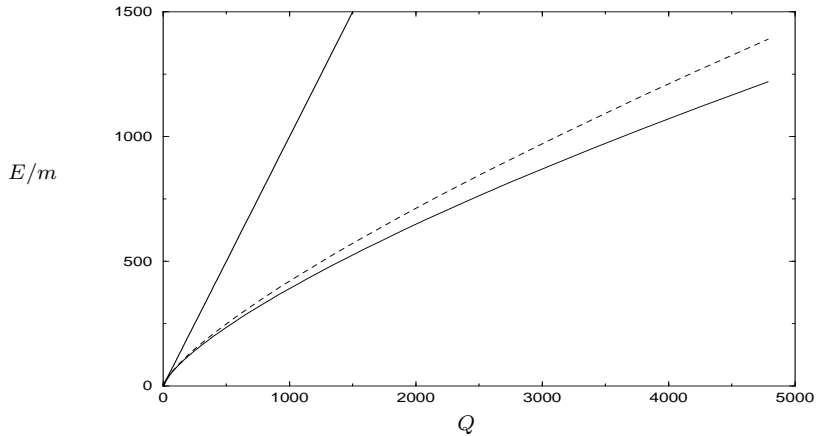}
\put(-10,325){$\scriptstyle Q$}
\put(-180,420){$\scriptstyle E/m$}
\end{picture}
\caption{The total energy $E_{\rm tot}/m$ (solid line) 
and the scalar field energy $E_{\rm s}/m$ (dotted line) as function of the
charge $Q$. The smallest $Q$ corresponds to $\omega/m=0.995$ whereas the largest one
corresponds to $\omega/m=0.261$. The straight line is the $E=mQ$ curve.}\label{kuva4}
\end{figure}

For numerical studies of potential (\ref{pot1}) we choose
$\lambda=1$, $M=0.0316\MP $ and $\epsilon = 10^{-3}$.
In Fig. \ref{kuva3}a we have plotted the field configuration 
corresponding to several different values of 
$\omega/m = 0.5,\ 0.4,\ 0.3,\ 0.261$.
The ratio $(B(r)-1)/(A(r)-1)$, which inside the bubble equals $(2-\kappa)/(1+\kappa)$, is presented
in Fig. \ref{kuva3}b. Moreover,
we now also present in Fig. \ref{kuva4} the energies $E_{\rm tot}$ and $E_{\rm s}$ 
as function of the charge $Q$. For these values of $m,\lambda$ and $\epsilon$ the 
critical $\omega$ is simply given by $\omega_c^2 = 10^{-3}m^2$. Note, 
that this time
the smallest $\omega$ we are able to calculate is $\omega = 0.261\, m$ which is
significantly bigger that the critical $\omega \simeq 0.0316\, m$.
This reflects the 
fact, that the thin-wall type behaviour begins at much larger ratio $\omega /\omega_c$
for a type II potential that for a type I potential. From Fig. \ref{kuva3}b one can also 
read that
the effective parameter $\kappa$ inside the bubble for small $\omega$ is 
about 0.07. The difference between total and scalar field energies, for the largest plotted 
balls $Q=4.8\times 10^3$, is about 13\%.

\section{Discussion and Conclusions}

In this paper we have considered the effects of gravity on Q-ball type
scalar field configurations. Including the effects of gravity leaves 
the flat space Q-ball \ansatz valid; one still has a spherically symmetric
solution whose complex phase rotates uniformly with time.
The geometry of space-time is modified from the flat case, which
gives a contribution to the charge and energy of the configuration.

From the analytical and numerical work it is evident that the effects 
of gravity can give a large contribution to the Q-ball energy.
However, in order for the contribution to be significant,
either the free quanta of the Q-ball scalar field need to be extremely heavy
or the charge of the Q-ball needs to be very large. The gravitational
contribution is strongly dependent on the details of the potential
and needs, in general, to be computed numerically. 

Interesting points are raised by the analytical analysis. The analytical 
work is concentrated on the
thin-wall limit, in order to gain an understanding of the properties
of large Q-balls for which gravitational effects are more likely to be
significant. In the thin-wall limit, analytical considerations reveal that
the maximal size of Q-balls is ultimately limited by gravity.
The non-trivial geometry inside the Q-ball limits the size of Q-balls
for any potential that allows for Q-ball solutions. The physical 
reasoning behind this is simple. In equilibrium, the rotation of the complex phase
generates an outward pressure which prevents the Q-ball
from collapsing due to gravitational forces. In the ordinary thin-wall limit,
the Q-ball frequency tends a critical value $\w_c$, while the 
charge the Q-ball becomes larger and larger. If gravity is included in the
scenario, it is evident that at some point the Q-ball becomes too large
to be balanced by the internal pressure. Hence, there exists another
critical value, $\w_G>\w_c$, corresponding to the maximally large Q-ball.

The limit on the maximal Q-ball size due to the gravitational effects
is an interesting point raised by this analysis. 
The two example potentials  discussed approximately correspond to
the potentials associated with gauge (type I) and gravity (type II) 
mediated supersymmetry breaking.
Using typical values for the variables, $m\sim 100\GeV$, $\varphi_1\sim M\sim 10^{15}\GeV$,
and approximating $\w_c\sim m$, estimates for the maximum charge, energy and radius are:
$Q_{max}\sim 10^{60},\ E_{\max}\sim 10^{49}\GeV,\ R_{max}\sim 10^{13}\GeV^{-1}\sim 0.1\cm$ (type I) and $Q_{max}\sim 10^{34},\ E_{max}\sim 10^{36}\GeV,\ R_{max}\sim 1\GeV^{-1}$ (type II). In both cases, the Schwarzschild radius of a corresponding black hole is still
much smaller than that of the Q-ball.
Gravity will clearly not have a very significant role 
to play in a cosmological setting for Q-balls in the gauge mediated scenario. 
In the gravity mediated scenario the limit is much lower and may be significant.
It is interesting to note that, in order for the primordial Q-balls
to act as a self-interacting dark matter candidate \cite{kusespe},
one typically has to consider Q-balls with very large charges \cite{siqdm}.

Since the gravitational contribution to the Q-ball energy is negative,
it is possible that gravity can render an otherwise energetically unstable
Q-ball stable. This behaviour was also confirmed by the numerical work,
where is was observed that the negative energy of the gravitational field 
could stabilise a large enough Q-ball.
One can then envision a cosmological scenario where Q-balls are unstable,
and hence do not exist, unless their charge is large enough for the
gravitational effects to stabilise them. A primordial Q-ball distribution
formed at a large energy scale would then leave behind only stable, super heavy
relic Q-balls.


\section*{Acknowledgements}
This work has been partly supported by the Magnus Ehrnrooth Foundation.

\end{document}